\documentclass[12pt]{article}
\usepackage{graphicx}

\newsavebox{\sboxpubnumber}
\newsavebox{\sboxpubdate}

\newcommand{\Title}[1]{\begin{center} {\Large #1 } \end{center}}
\newcommand{\Author}[1]{\begin{center}{ \sc #1} \end{center}}
\newcommand{\Address}[1]{\begin{center}{ \it #1} \end{center}}

\newenvironment{Abstract}{\begin{quotation}  }{\end{quotation}}
\newenvironment{Presented}{\begin{quotation} \begin{center}
             PRESENTED AT\end{center}\bigskip
      \begin{center}\begin{large}}{\end{large}\end{center}
      \end{quotation}}

\def\ds{\displaystyle}
\def\lesssim{{\
\lower-1.2pt\vbox{\hbox{\rlap{$<$}\lower5pt\vbox{\hbox{$\sim$}}}}\ }} 
\def\gtrsim{{\
\lower-1.2pt\vbox{\hbox{\rlap{$>$}\lower5pt\vbox{\hbox{$\sim$}}}}\ }}

\begin{document}

\begin{titlepage}

\vfill
\Title{Affleck-Dine baryogenesis and the Q-ball dark matter \\
in the gauge-mediated SUSY breaking}
\vfill
\Author{S. Kasuya}

\Address{Research Center for the Early Universe, University of Tokyo \\
         Bunkyo-ku, Tokyo 113-0033, Japan}
\vfill
\begin{Abstract}
We consider the Affleck-Dine baryogenesis comprehensively in the
minimal supersymmetric standard model with gauge-mediated
supersymmetry breaking. Considering the high temperature effects, we
see that the Affleck-Dine field is naturally deformed into the form
of the Q ball. In the natural scenario where the initial amplitude of
the field and the A-terms are both determined by the nonrenormalizable
superpotential, we obtain a narrow allowed region in the
parameter space in order to explain the baryon number and the dark
matter of the universe simultaneously. Therefore, the Affleck-Dine
baryogenesis is successful, although difficult.
\end{Abstract}
\vfill
\begin{Presented}
    COSMO-01 \\
    Rovaniemi, Finland, \\
    August 29 -- September 4, 2001
\end{Presented}
\vfill
\end{titlepage}
\def\thefootnote{\fnsymbol{footnote}}
\setcounter{footnote}{0}

\section{Introduction}
The Affleck-Dine (AD) mechanism \cite{AfDi} is the most promising
scenario for explaining the baryon number of the universe. It is based
on the dynamics of a (complex) scalar field $\phi$ carrying baryon 
number, which is called the AD field. During inflation, the
expectation value of the AD field develops at a very large value. 
After inflation the inflaton field oscillates about the minimum of the
effective potential, and dominates the energy density of the universe
like matter, while the AD field stays at the large field value. It 
starts the oscillation, or more precisely, rotation in its effective
potential when $H \sim m_{\phi,eff}$, where $H$ and 
$m_{\phi,eff}\equiv|V''(\phi)|$ are the
Hubble parameter and the curvature of the potential of the AD
field. Once it rotates, the baryon number will be created as 
$n_{B} \sim \omega \phi^2$, where $\omega = \dot{\theta}$ is the
velocity of the phase of the AD field. When $t \sim \Gamma_{\phi}$,
where $\Gamma_{\phi}$ is the decay rate of the AD field, it decays
into ordinary particles carrying baryon number such as quarks, and the 
baryogenesis in the universe completes.

However, important effects on the field dynamics were overlooked. It
was recently revealed that the AD field feels spatial instabilities
\cite{KuSh}. Those instabilities grow very large and the AD field
deforms into clumpy objects: Q balls. A Q ball is a kind of
nontopological soliton, whose stability is guaranteed by the existence
of some charge $Q$ \cite{Coleman}. In the previous work \cite{KK1}, we
found that all the charges which are carried by the AD field are
absorbed into formed Q balls, and this implies that the baryon number
of the universe cannot be explained by the relic AD field remaining
outside Q balls after their formation.

In the radiation dominated universe, charges are evaporated from Q
balls \cite{LaSh}, and they will explain the baryon number of the
universe \cite{KK3}. This is because the minimum of the (free) energy
is achieved when the AD particles are freely in the thermal plasma at
finite temperature. (Of course, the mixture of the Q-ball
configuration and free particles is the minimum of the free
energy at finite temperature, when the chemical potential of the Q
ball and the plasma are equal to be in the chemical equilibrium. This
situation can be achieved for very large charge of Q balls with more
than $10^{40}$ or so \cite{LaSh}.) Even if the radiation component is
not dominant energy in the universe, such as that during the
inflaton-oscillation dominant stage just after the inflation, high
temperature effects on the dynamics of the AD field and/or Q-ball
evaporation are important. Therefore, in this article, we investigate
the whole scenario of the AD mechanism for baryogenesis in the minimal
supersymmetric standard model (MSSM) with the gauge-mediated
supersymmetry (SUSY) breaking in the high temperature universe.

\section{Flat directions as the Affleck-Dine field}
A flat direction is the direction in which the effective potential
vanishes. There are many flat directions in the minimal supersymmetric 
standard model (MSSM), and they are listed in
Refs.~\cite{DiRaTh,GhKoMa}. Since they consist of squarks and/or
sleptons, they carry baryon and/or lepton numbers, and we can identify 
them as the Affleck-Dine (AD) field. Although the flat directions are
exactly flat when supersymmetry (SUSY) is unbroken, it will be lifted
by SUSY breaking effects. In the gauge-mediated SUSY breaking
scenario, SUSY breaking effects appear at low energy scales, so the
shape of the effective potential for the flat direction has curvature
of order of the electroweak mass at low scales, and almost flat at
larger scales. Therefore, the effective potential reads as
\begin{eqnarray}
    \label{pot-1}
    V(\Phi) & = & 
       M_F^4 \log \left( 1+\frac{|\Phi|^2}{M_S^2} \right)
     + m_{3/2}^2 \left[ 1+K\log\left(\frac{|\Phi|^2}{M^2}
                             \right)\right]|\Phi|^2
     \nonumber \\
     & & + \lambda^2\frac{|\Phi|^{2n-2}}{M^{2n-6}}
         + \left( \lambda A_{\lambda} \frac{\Phi^n}{M^{n-3}} 
                  + h.c. \right)
     \nonumber \\
     & & - c_H H^2|\Phi|^2
         + \left( \lambda a_H H \frac{\Phi^n}{M^{n-3}} 
                  + h.c. \right)
     \nonumber \\
     & & + c_T^{(1)} T^2 |\Phi|^2
         + c_T^{(2)} T^4 \log \frac{|\Phi|^2}{T^2},
\end{eqnarray}
where $\Phi$ is the complex scalar field representing the flat
direction, $M_S \sim M_F^2/m_{\phi}$ is a messenger scale. The second
term comes from the gravity mediation effects, since the gravity
effects always exist.  The $K$ is the coefficient of the the one-loop
corrections \cite{EnMc1}, and is usually negative. However it may be
positive in some cases, and Q balls are not formed until the amplitude
of the field becomes as small as it stays in the logarithmic
potential. We call it {\it delayed} case. $M=2.4\times 10^{18}$ GeV is
the reduced Planck mass. The second line represents for the
nonrenormalizable terms. The third line shows terms which depends on
the Hubble parameter $H$, during inflation and inflaton oscillation
stage which starts just after inflation. $c_H$ is a positive constant
and $a_H$ is a complex constant with a different phase from
$A_{\lambda}$ in order for the AD mechanism to work. The last line is
the finite temperature potential. The first and second term denote
direct and indirect coupling of the flat direction to thermal bath
\cite{AnDi,FuHaYa}, respectively. 

\section{Affleck-Dine mechanism}
\label{AD-mech}
It was believed that the Affleck-Dine (AD) mechanism works as follows
\cite{AfDi,DiRaTh}. During inflation, the AD field are trapped at the
value determined by the following conditions: 
$V_H'(\Phi) \sim V_{NR}'(\Phi)$ and $V_{AH}'(\Phi) \sim 0$. Therefore, 
we get 
\begin{eqnarray}
    \label{AD-min}
    & & \phi_{min} \sim (HM^{n-3})^{1/(n-2)}, \nonumber \\
    & & \sin[n\theta_{min}+arg(a_H)] \sim 0,
\end{eqnarray}
where $c_H,\lambda \sim 1$ are assumed, and 
$\Phi=(\phi e^{i \theta})/\sqrt{2}$. After inflation, the inflaton
oscillates around one of the minimum of its effective potential, and
the energy of its oscillation dominates the universe. During that
time, the minimum of the potential of the AD field are adiabatically
changing until it rolls down rapidly when $H \simeq \omega\equiv
|V'(\phi)/\phi|^{1/2}$. Then the AD field rotates in its potential,
producing the baryon number $n_B = \dot{\theta} \phi^2$. After the AD
field decays into quarks, baryogenesis completes. 

In order to estimate the amount of the produced baryon number, let us
assume that the phases of $A_{\lambda}$ and $a_H$ differ of order
unity. Then the baryon number can be estimated as
\begin{equation}
    n_B  \sim  H^{-1} \frac{\partial V_A}{\partial \phi}\phi 
          \sim H^{-1} V_A 
     \sim \omega^{-1} m_{3/2} \frac{\phi^n}{M^{n-3}}
      \sim   \left( \frac{m_{3/2}}{\omega} \right) \omega\phi^2
      = \varepsilon n_B^{(max)},
\end{equation}
where $\varepsilon \equiv (m_{3/2}/\omega)$, 
$n_B^{(max)} \equiv \omega\phi^2$, and $H \sim \omega$ is used in the
second line. Notice that the contribution from the Hubble A-term is at 
most comparable to this. When $\varepsilon=1$, the trace of the motion
of the AD field in the potential is circular orbit. If $\varepsilon$
becomes smaller, the orbit becomes elliptic, and finally the field is
just oscillating along radial direction when $\varepsilon=0$. We call
$\varepsilon$ as the ellipticity parameter below.

When the logarithmic potential is dominant, $\omega \sim m^2/\phi$, so
the ellipticity parameter varies, and it may be very small. On the
other hand, when the potential is dominated by the gravity effect, 
$\omega\simeq m_{3/2}$. Then $\varepsilon \simeq 1$ always holds.

\section{Charge evaporation at high temperature}
\subsection{Gauge-mediation type Q balls}
Since the Q-ball formation takes place nonadiabatically, (almost) all
the charges are absorbed into produced Q balls \cite{KK1,KK2}. Thus,
the baryon number of the universe cannot be explained by the remaining
charges outside Q balls in the form of the relic AD field. However, at
finite temperature, the minimum of the free energy of the AD field is
achieved for the situation that all the charges exist in the form of
free particles in thermal plasma \cite{LaSh}. This situation is
realized through charge evaporation from Q-ball surface
\cite{LaSh,EnMc2}. In spite of the fact that the complete evaporation
is the minimum of the free energy, the actual universe is filled with
the mixture of Q balls and surrounding free particles, since the
evaporation rate becomes smaller than the cosmic expansion rate at low 
temperatures. (As we mentioned in the Introduction, the free energy is 
minimized at the situation that some charges exist in the form of free
particles (in thermal plasma) and the rest stays inside Q balls, if
the Q-ball charge are large enough for the chemical equilibrium
\cite{LaSh}. In this case, the charge of the Q ball should be 
$Q \sim \eta_B^{-4} \sim 10^{40}$.) The nonadiabatic creation of Q
balls and the later charge evaporation takes place, since the time
scale of evaporation is much longer than that of the Q-ball 
formation. Notice that the charge of the Q ball is conserved for the
situation that the evaporation of charges is not effective. However,
the energy of the Q ball decreases as the temperature of the universe
decreases. This happens because the Q ball and surrounding plasma are
in thermal equilibrium: i.e., the Q balls and thermal plasma have the
same temperatures.

The rate of charge transfer from inside the Q ball to its outside is
determined by the diffusion rate at high temperature and the
evaporation rate at low temperature \cite{KK3}. When the
difference between the chemical potentials of the plasma and the Q
ball is small, chemical equilibrium is achieved and charges inside the
Q ball cannot come out \cite{BaJe}. Therefore, the charges in the
`atmosphere' of the Q ball should be taken away in order for further
charge evaporation. This process is determined by the diffusion. The
diffusion rate is given by \cite{BaJe}
\begin{equation}
    \label{diff-rate}
    \Gamma_{diff} \equiv \frac{dQ}{dt} \sim -4\pi D R_Q \mu_Q T^2
                                       \sim -4\pi A T, 
\end{equation}
where $D=A/T$ is the diffusion coefficient, and $A=4-6$, $\mu_Q \sim 
\omega$ is the chemical potential of the Q ball. On the other hand,
the evaporation rate is \cite{LaSh}
\begin{equation}
    \label{evap-rate}
    \Gamma_{evap} \equiv \frac{dQ}{dt} 
     \sim  -\zeta(\mu_Q-\mu_{plasma})T^2 4\pi R_Q^2
     \sim  -4\pi\zeta \frac{T^2}{M_F}Q^{1/4},
\end{equation}
where $\mu_{plasma}\ll \mu_Q$ is the chemical potential of thermal
plasma, and $\mu_Q \sim \omega \sim M_F Q^{-1/4}$ is used in the
second line. $\zeta$ change at $T=m_{\phi}$ as
\begin{equation}
    \zeta=\left\{
      \begin{array}{cl}
          \ds{\left(\frac{T}{m_{\phi}}\right)^2} 
                    & \ds{(T<m_{\phi}),} \\[2mm]
          \ds{1}  & \ds{(T>m_{\phi}).}
      \end{array}
      \right.
\end{equation}
Therefore, we get 
\begin{equation}
    \label{evap-rate-2}
    \Gamma_{evap} = \frac{dQ}{dt} = \left\{
      \begin{array}{ll}
          \ds{-4\pi \frac{T^2}{M_F} Q^{1/4}} & \ds{(T>m_{\phi}),} \\
          \ds{-4\pi \frac{T^4}{m_{\phi}^2 M_F}Q^{1/4}} & 
                      \ds{(T<m_{\phi}).} 
      \end{array}
      \right.
\end{equation}

Integrating the above expressions, we finally obtain the total
evaporated charge as
\begin{equation}
    \label{evap-Q}
    \Delta Q \sim 4.6 \times 10^{17} 
              \left(\frac{m_{\phi}}{{\rm TeV}}\right)^{-2/3}
              \left(\frac{M_F}{10^6 {\rm GeV}}\right)^{-1/3}
              \left(\frac{Q}{10^{24}}\right)^{1/12}.
\end{equation}

\subsection{Gravity-mediation type Q balls}
Now we will show the evaporated charges for the `new' type of stable Q
ball \cite{KK3}. The evaporation and diffusion rate have the same
forms in terms of Q-ball parameters $R_Q$ and $\omega$. The only
differences are that we have to use the features for the
`gravity-mediation' type Q ball, such as 
\begin{equation}
    R_Q \sim |K|^{-1/2} m_{3/2}, \qquad \omega \sim m_{3/2},
\end{equation}
and the transition temperature when $\Gamma_{evap}=\Gamma_{diff}$
becomes $T_*\equiv A^{1/3}|K|^{1/6}(m_{3/2}m_{\phi}^2)^{1/3}$. As in
the `usual' type of Q balls, where the potential is dominated by the
logarithmic term, the charge evaporation near $T_*$ is dominant, and
the total evaporated charges are found to be \cite{KK3}
\begin{equation}
    \Delta Q \sim 10^{20} \left(\frac{m_{3/2}}{\rm MeV}\right)^{-1/3}
       \left(\frac{m_{\phi}}{\rm TeV}\right)^{-2/3}.
\end{equation}

\section{Q-ball formation}
Q-balls are produced while the AD field is rotating. We obtained the
relation between the typical charge of the Q balls and the initial
amplitude of the AD field \cite{KK6}. Here we only show the results.
For the gauge-mediation type of Q balls, we have
\begin{equation}
    Q = \beta \left( \frac{\phi_0}{m(T)} \right)^4,
\end{equation}
where $\beta \simeq 6 \times 10^{-4}$, and $m(T)=M_F$ for $T<M_F$,
while $m(T)=T$ for $T>M_F$. For the gravity-mediation type, 
\begin{equation}
    Q = \beta' \left( \frac{\phi_0}{m_{3/2}} \right)^2,
\end{equation}
where $\beta' \simeq 6\times 10^{-3}$. These are consistent with the
analytical estimation: $Q \sim k_{res}^{-3} n_b,$ where $k_{res}^{-1}$
is the size of the resonance mode, and  $n_b \sim \omega\phi_0^2$ is
the baryon density.

\section{General constraints for Q-ball scenarios}
We would like to see whether there is any consistent cosmological
scenario for the baryogenesis and the dark matter of the universe,
provided by large Q balls. In the Q-ball scenario, the baryon number
of the universe should be explained by the amount of the charge
evaporated from Q balls, $\Delta Q$, and the survived Q balls become
the dark matter. If we assume that Q balls do not exceed the critical
density of the universe, i.e., $\Omega_Q \lesssim 1$, and the
baryon-to-photon ratio as $\eta_B \sim 10^{-10}$, the condition can be
written as
\begin{equation}
    \label{eta-B}
    \eta_B = \frac{n_B}{n_{\gamma}} 
       \simeq \frac{\varepsilon n_Q\Delta Q}{n_{\gamma}}
       \simeq \frac{\varepsilon\rho_Q \Delta Q}{n_{\gamma}M_Q}
       \simeq \frac{\varepsilon\rho_{c,0} \Omega_Q \Delta Q}
                   {n_{\gamma,0}M_Q},
\end{equation}
where $\Omega_Q$ is the density parameter for the Q ball. Since the
evaporated charge explains the baryon number in the universe, while
survived Q-balls becomes the dark matter, we can have the Q-ball
charge to explain the right amount of both baryons and dark matter in
the universe, using $M_Q \simeq m_{\phi}Q^{3/4}$, 
$\rho_{c,0} \sim 8h_0^2\times 10^{-47} {\rm GeV}^4$, and
$n_{\gamma,0} \sim 3.3 \times 10^{-39} {\rm GeV}^3$, where 
$h_0(\sim 0.7)$ is the Hubble parameter normalized with 100km/sec/Mpc.
It reads as
\begin{equation}
    Q \sim 3.2\times 10^{17} \varepsilon^{3/2} \Omega_Q^{3/2}
    \left(\frac{\eta_B}{10^{-10}}\right)^{-3/2}
    \left(\frac{m_{\phi}}{{\rm TeV}}\right)^{-1}
    \left(\frac{M_F}{10^6 {\rm GeV}}\right)^{-2}.
\end{equation}
We call this baryon-dark matter (B-DM) condition. We can also put
another condition, namely. survival condition, which implies that
the Q ball should survive from the evaporation: $Q \gtrsim \Delta Q$.
It is expressed as
\begin{equation}
    \label{survive-2}
    Q \gtrsim 1.2\times 10^{17} 
      \left(\frac{m_{\phi}}{\rm TeV}\right)^{-8/11}
      \left(\frac{M_F}{10^6 {\rm GeV}}\right)^{-4/11}.
\end{equation}
Finally, Q balls must be stable against the decay into nucleons to
become the dark matter of the universe, which we call it the stability 
condition:
\begin{equation}
    Q \gtrsim 10^{24} \left(\frac{M_F}{10^6 {\rm GeV}}\right)^4.
\end{equation}

\section{Consistent cosmological Q-ball scenarios}
\begin{figure}[htb]
\centering
\includegraphics[height=3.7in]{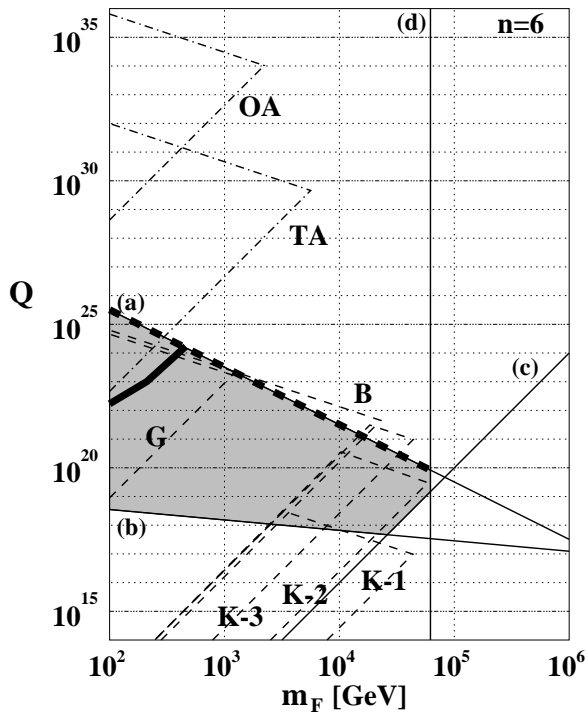}
\caption{
Summary of constraints on the parameter space ($Q,M_F$) with $n=6$ and 
$m_{\phi}=100$ GeV in the generic logarithmic potential where the
thermal terms are dominated when the Q-ball formation occurs. 
Lines (a), (b), and (c) denote the B-DM, survival, and stability
conditions, respectively. Line (d) shows the conditions 
$T \gtrsim M_F$. }
\label{fig1} 
\end{figure}

Now we can see whether there is any consistent scenario to explain the 
amounts of baryons and dark matter of the universe naturally, which
means that the initial conditions and the A-terms are provided by the
non-renormalizable superpotential. The details can be seen in
Ref.\cite{KK6}, so we will summarize the results here. There are only
three successful cases. The first one is when the potential is
dominated by the thermal logarithmic term, which is shown in
Fig.~\ref{fig1}. Here, the successful scenario is achieved if the
parameters are as follows: $M_F\sim 5\times 10^2$ GeV, $T_{RH} \sim 5
\times 10^4$ GeV, and $Q \sim 2 \times 10^{24}$.

\begin{figure}[htb]
\centering
\includegraphics[height=3.7in]{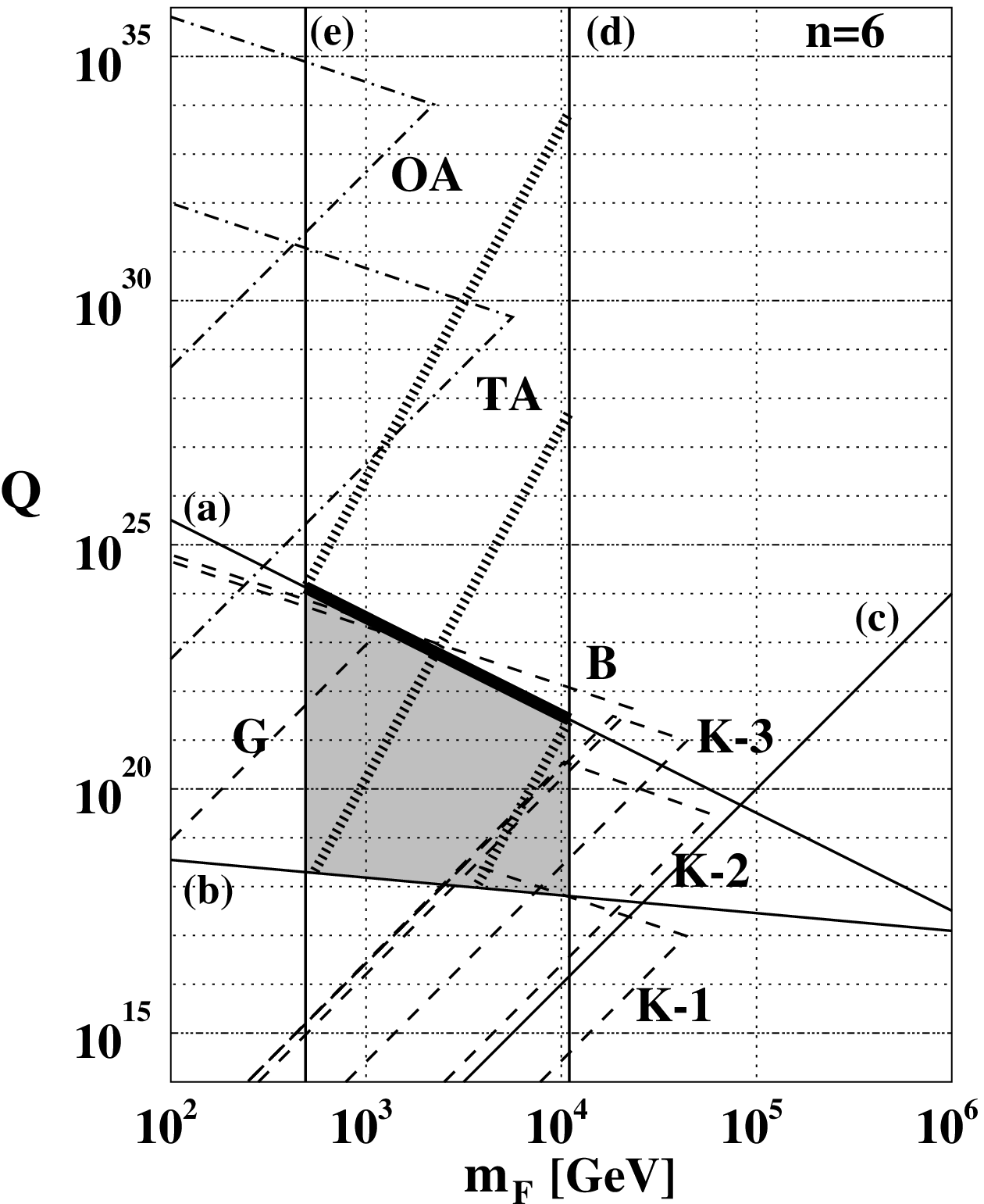}
\caption{
Summary of constraints on the parameter space $(Q,M_F)$ for the
delayed-formed Q balls with the thermal logarithmic term domination
for $m_{\phi}=100$ GeV and $n=6$. Lines (a), (b), and (c) denote the
B-DM, survival, and stability conditions, respectively. Lines (d) and
(e) show the conditions $T_{eq} \gtrsim M_F$ and $m_{3/2} \lesssim 1$
GeV, respectively. }
\label{fig2} 
\end{figure}

The second case is the delayed case with the Q-balls formed when the
thermal logarithmic term is dominated, shown in Fig.~\ref{fig2}. In
this case, the allowed parameter sets are $m_{3/2} \sim $ GeV, $M_F
\sim 5 \times 10^2$ GeV, $T_{RH} \sim 10^7$ GeV, and $Q \sim 2 \times
10^{24}$.

\begin{figure}[htb]
\centering
\includegraphics[height=3.7in]{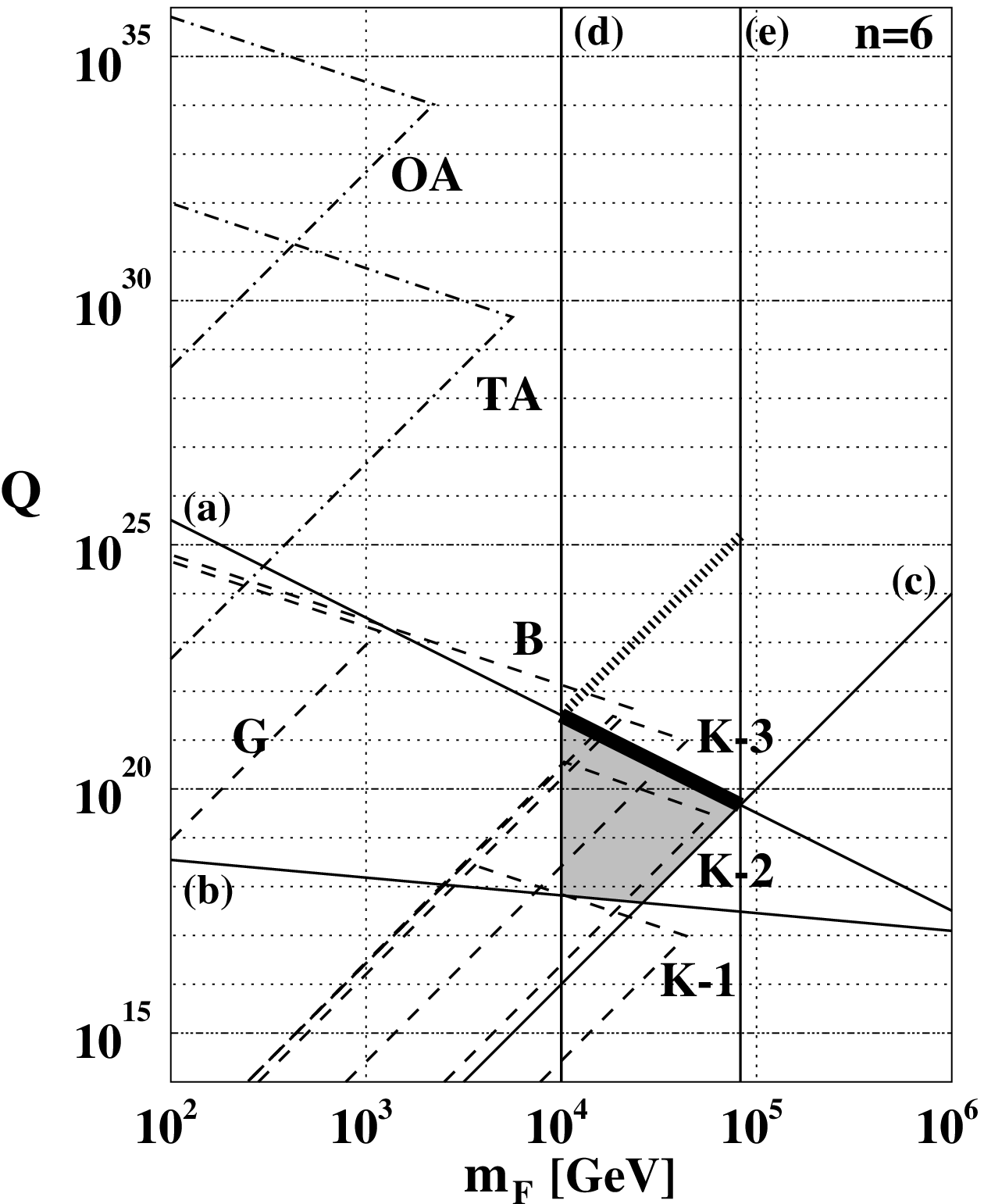}
\caption{
Summary of constraints on the parameter space $(Q,M_F)$ for the
delayed-formed Q balls with the generic logarithmic term domination
for $n=6$ and $m_{\phi}=100$ GeV. Lines (a), (b), and (c) denote the
B-DM, survival, and stability conditions, respectively. Line (d)
represents the condition  $M_F\gtrsim T_{eq}$, and line (e) is just
the upper limit for $M_F$.} 
\label{fig3} 
\end{figure}

The last case is the delayed case with the Q-balls formed when the
zero-temperature logarithmic term is dominated, shown in
Fig.~\ref{fig3}. In this case, the allowed parameter sets are $m_{3/2}
\sim 0.1$ GeV, $M_F \sim 5 \times 10^4$ GeV, $T_{RH} \sim 5$ GeV, and
$Q \sim 10^{20}$.

As can be seen, we need very low reheating temperature and very low
$M_F$, which represents for the expectation value of the F-term of the 
messenger sector, for all the successful scenarios. Therefore, the
model is constrained rather severely in order to explain the baryons
and the dark matter of the universe in these Q-ball scenarios.

\section{Conclusion}
We have investigated thoroughly the Q-ball cosmology (the baryogenesis
and the dark matter) in the gauge-mediated SUSY breaking model. Taking 
into account thermal effects, the shape of the effective potential has 
to be altered somewhat, but most of the features of the Q-ball
formation derived at zero temperature can be applied to the finite
temperature case with appropriate rescalings. We have thus found that
Q balls are actually formed through Affleck-Dine mechanism in the
early universe. 

We have sought for the consistent scenario for the dark matter Q ball,
which also provides the baryon number of the universe
simultaneously. For the consistent scenario, we adopt the
nonrenormalizable superpotential in order to naturally give the
initial amplitude of the AD field and the source for the field
rotation due to the A-term. As opposed to our expectation, very narrow
parameter region could be useful for the scenario in the situations
that the Q balls are produced just after the baryon number
creation. In addition, we have seen that current experiments have
already excluded most of the successful parameter regions. 

Of course, if the A-terms and/or the initial amplitude of the AD field 
are determined by other mechanism, the cosmological Q-ball scenario
may work. Then, the stable dark matter Q balls supplying the baryons
play a crucial role in the universe. 

We have also found the new situations that the Q-ball formation takes
place when the amplitude of the fields becomes small enough to be in
the logarithmic terms in the potential, while the fields starts its
rotation at larger amplitudes where the effective potential is
dominated by the gravity-mediation term with {\it positive} K-term. 
This allows to produce Q balls with smaller charges while creating
larger baryon numbers. In this situation, there is wider allowed
regions for naturally consistent scenario, although the current
experiments exclude most of the parameter space. Notice also that
rather too low reheating temperature is necessary for larger $n$
scenario to work naturally. This aspect is good for evading the
cosmological gravitino problem, while it is difficult to construct the
actual inflation mechanism to get such low reheating temperatures, in
spite of the fact that the nucleosynthesis can take place successfully
for the reheating temperature higher than at least 10 MeV.

\end{document}